\documentclass[aps,prd,nofootinbib,amsmath,amssymb,superscriptaddress,showpacs,showkeys,twocolumn,10pt]{revtex4}%superscriptaddress,showpacs,

\pdfoutput=1

\usepackage{graphicx}
\usepackage{dcolumn}
\usepackage{bm}
\usepackage{amssymb}
\usepackage{latexsym}
\usepackage{booktabs}
\usepackage{amsmath}
\usepackage{multirow}
\usepackage[colorlinks=true, linkcolor=red, citecolor=blue]{hyperref}

\newcommand{\be}{\begin{equation}}
\newcommand{\ee}{\end{equation}}
\newcommand{\bq}{\begin{eqnarray}}
\newcommand{\eq}{\end{eqnarray}}

\usepackage[usenames,dvipsnames]{xcolor}

\DeclareMathAlphabet\mathbfcal{OMS}{cmsy}{b}{n}
\definecolor{darkgreen}{cmyk}{0.85,0.2,1.00,0.2}
\definecolor{purple}{cmyk}{0.5,1.0,0,0}

\def\barray{\begin{array}}
\def\earray{\end{array}}
\def\be{\begin{equation}}
\def\ee{\end{equation}}
\def\ben{\begin{equation} \nonumber}
\def\een{\end{equation}}
\def\ban{\begin{eqnarray*}}
\def\ean{\end{eqnarray*}}
\def\ba{\begin{eqnarray}}
\def\ea{\end{eqnarray}}

\def\({\left(}
\def\){\right)}

\begin{document}

\title{Searching for sterile neutrinos in dynamical dark energy cosmologies}

\author{Lu Feng}
%\email{fengluu@foxmail.com}
\affiliation{Department of Physics, College of Sciences, Northeastern University, Shenyang
110004, China}
\author{Jing-Fei Zhang}
%\email{jfzhang@mail.neu.edu.cn}
\affiliation{Department of Physics, College of Sciences, Northeastern University, Shenyang
110004, China}
\author{Xin Zhang\footnote{Corresponding author}}
\email{zhangxin@mail.neu.edu.cn}
\affiliation{Department of Physics, College of Sciences, Northeastern University, Shenyang 110004, China}
\affiliation{Center for High Energy Physics, Peking University, Beijing 100080, China}

\begin{abstract}

We investigate how the dark energy properties change the cosmological limits on sterile neutrino parameters by using recent cosmological observations. We consider the simplest dynamical dark energy models, the $w$CDM model and the holographic dark energy (HDE) model, to make an analysis. The cosmological observations used in this work include the Planck 2015 CMB temperature and polarization data, the baryon acoustic oscillation data, the type Ia supernova data, the Hubble constant direct measurement data, and the Planck CMB lensing data. We find that, $m_{\nu,{\rm{sterile}}}^{\rm{eff}}<0.2675$ eV and $N_{\rm eff}<3.5718$ for $\Lambda$CDM cosmology, $m_{\nu,{\rm{sterile}}}^{\rm{eff}}<0.5313$ eV and $N_{\rm eff}<3.5008$ for $w$CDM cosmology, and $m_{\nu,{\rm{sterile}}}^{\rm{eff}}<0.1989$ eV and $N_{\rm eff}<3.6701$ for HDE cosmology, from the constraints of the combination of these data. Thus, without the addition of measurements of growth of structure, only upper limits on both $m_{\nu,{\rm{sterile}}}^{\rm{eff}}$ and  $N_{\rm eff}$ can be derived, indicating that no evidence of the existence of a sterile neutrino species with eV-scale mass is found in this analysis. Moreover, compared to the $\Lambda$CDM model, in the $w$CDM model the limit on $m_{\nu,{\rm{sterile}}}^{\rm{eff}}$ becomes much looser, but in the HDE model the limit becomes much tighter. Therefore, the dark energy properties could significantly influence the constraint limits of sterile neutrino parameters.

\end{abstract}

\maketitle

\section{Introduction}
\label{sec1}
Neutrino oscillation experiments with solar and atmospheric neutrinos have already provided evidence for the non-zero neutrino mass \cite{Lesgourgues:2006nd} (see Refs.~\cite{Zhang:2017rbg,Zhao:2016ecj,Wang:2016tsz,Huang:2015wrx,Zhang:2015uhk,Zhang:2015rha,Chen:2015oga,Wang:2012uf,Drewes:2013gca,GonzalezGarcia:2007ib,Guo:2017hea,Capozzi:2017ipn,Li:2017iur,Vagnozzi:2017ovm,Lu:2016hsd,Capozzi:2016rtj,DellOro:2015kys,Giusarma:2016phn,DiValentino:2015ola,DiValentino:2016hlg,Yang:2017amu,Dai:2017sst,Zhu:2014qma,Xu:2016ddc} for the studies of the total neutrino mass in cosmology). There is still room for extra sterile species, supported by anomalies in short-baseline and reactor neutrino experiments~\cite{GonzalezGarcia:2012sz,Gariazzo:2013gua,Giunti:2013aea,Kopp:2013vaa,Giunti:2012bc,Giunti:2012tn,Aguilar-Arevalo:2012fmn,Conrad:2012qt,Mention:2011rk,Giunti:2010zu,Aguilar:2001ty,Aguilar-Arevalo:2013pmq,Conrad:2013mka}. It seems that the fully thermalized ($\Delta N_{\rm eff}\approx1$) sterile neutrinos with eV-scale mass are needed to explain these anomaly results \cite{Abazajian:2012ys,Hannestad:2012ky,Conrad:2013mka}.

However, recently, the neutrino oscillation experiments by the Daya Bay and MINOS collaborations \cite{Adamson:2016jku}, as well as the cosmic ray experiment by the IceCube collaboration \cite{TheIceCube:2016oqi}, found no evidence for a sterile neutrino species with eV-scale mass, in tension with the previous short-baseline neutrino oscillation experiments that prefer a sterile neutrino species with mass around 1 eV. Therefore, it is particularly important to search for cosmological evidence of sterile neutrinos through their gravitational effects on the large-scale structure formation and the evolution of the universe.

In Ref.~\cite{Feng:2017nss}, it was shown that, in the $\Lambda$ cold dark matter ($\Lambda$CDM) model, a search for massive sterile neutrinos with the latest observations (Planck TT,TE,EE+lowP+BAO+$H_0$+SZ+lensing+WL) indicates that the mass of sterile neutrino (if it exists) is actually very light. Besides, involving a sterile neutrino species in the $\Lambda$CDM model can also relieve the tension between observations to some extent (see also Refs.~\cite{Zhang:2014dxk,Dvorkin:2014lea,Hamann:2013iba,Wyman:2013lza,Battye:2013xqa}) and improve the cosmological fit. For the related works about sterile neutrinos, see, e.g., Refs.~\cite{Zhao:2017urm,Zhang:2014ifa,Zhang:2014nta,Ko:2014bka,Archidiacono:2014apa,Archidiacono:2014nda,Li:2014dja,An:2014bik,Zhang:2014lfa,Li:2015poa,Palazzo:2013me,Abazajian:2017tcc,Gariazzo:2015rra,Gariazzo:2014pja,Kirilova:2014ipa,Girardi:2014wea,Luo:2014vha,Abazajian:2012ys}.

The investigation in Ref.~\cite{Feng:2017nss} for searching for sterile neutrinos is based on the $\Lambda$CDM cosmology. It is known that, though the cosmological constant $\Lambda$ \cite{Einstein:1917ce} is the simplest candidate for dark energy and actually the $\Lambda$CDM model can explain the various cosmological observations quite well \cite{Li:2009jx,Xu:2016grp}, it suffers from the difficult theoretical puzzles, such as the ``fine-tuning" and ``cosmic coincidence'' problems \cite{Weinberg:1988cp,Sahni:1999gb,Frieman:2008sn}. On the other hand, actually, many other dark energy candidates are not yet excluded by the current observational data \cite{Xu:2016grp,Li:2009jx}. In fact, some dynamical dark energy models are still rather competitive in fitting the current observations \cite{Xu:2016grp}. Therefore, it is important to investigate how a dynamical dark energy impacts the measurements of sterile neutrino parameters with the latest cosmological observations, compared to the case of $\Lambda$CDM cosmology.

In this paper, we will investigate how the dark energy properties could influence the cosmological constraints on the sterile neutrino parameters. There are still too many dark energy models that are not excluded by the observations, and thus it is of course not possible and not necessary to discuss them one by one in this paper. In Refs.~\cite{Zhang:2015uhk,Wang:2016tsz}, the issue of constraining the active neutrino mass in dynamical dark energy models was discussed, and the scheme was to choose some typical dark energy models as the simplest extensions to the $\Lambda$CDM cosmology to detect the effects of dark energy properties on constraining neutrino mass. In this paper, we adopt the same scheme to investigate the issue concerning the sterile neutrino.

Following Refs.~\cite{Zhang:2015uhk,Wang:2016tsz}, we also only consider the simplest dynamical dark energy models, i.e., we only choose the models that have only one more parameter compared to $\Lambda$CDM. One is the so-called $w$CDM model in which a constant equation-of-state parameter (EoS) $w$ is endowed for dark energy. Such a model is considered to be the simplest extension to the $\Lambda$CDM cosmology, but the drawback is obvious, i.e., there is no reason to remain $w$ as a constant in the actual physical consideration. A usual remedy is to consider a time-varying EoS by adopting the parametrization  $w(a)=w_0+w_a(1-a)$, but this introduces one more parameter and thus we will not consider this case in this paper. The other one we choose is the holographic dark energy (HDE) model \cite{Li:2004rb} (for a recent review, see Ref.~\cite{Wang:2016och}) that has only one additional dimensionless parameter, $c$, solely determining the cosmological evolution of dark energy in this model (note that here $c$ is not the speed of light, and we actually use the natural units in which the speed of light is equal to one). The HDE model originates from the holographic principle of quantum gravity theory. From some theoretical considerations, the effective quantum field theory combined with the requirement of the holographic principle of quantum gravity leads to the HDE model with the energy density of dark energy given by $\rho_{\rm de}=3c^2M_{\rm pl}^2 R_{\rm eh}^{-2}$, where $M_{\rm pl}$ is the reduced Planck mass and $R_{\rm eh}$ is the event horizon of the universe. The HDE model is expected to provide some clues for a bottom-up exploration of a quantum theory of gravity, and thus this model has been attracting extensive theoretical interests. In the HDE model, the evolution of EoS is given by $w(a)=-1/3-(2/3c)\sqrt{\Omega_{\rm de}(a)}$, where $\Omega_{\rm de}(a)$ is the solution of a differential equation. For the details of the evolution of dark energy in the HDE model, see the equations (2.4)--(2.7) in Ref.~\cite{Zhang:2015rha}. The HDE model has been widely studied (see, e.g., Refs.~\cite{Huang:2004ai,Zhang:2005yz,Zhang:2005hs,Chang:2005ph,Zhang:2006av,Zhang:2006qu,Zhang:2007sh,Zhang:2007uh,Zhang:2007es,Zhang:2007an,Ma:2007av,Zhang:2009un,Zhang:2009xj,Li:2009zs,Zhang:2012sya,Li:2013dha,Zhang:2014ija,Cui:2015oda,Wang:2013zca,He:2016rvp,Cui:2017idf}). In particular, it was shown in a recent study \cite{Xu:2016grp} on the comparison of popular dark energy models that the HDE model is still a competitive candidate of dark energy among many theoretical models.

This paper is organized as follows. In Sec. \ref{sec2}, we will describe the analysis method and the observational data we use in this paper. In Sec. \ref{sec3}, we present the fit results and discuss these results in detail. Conclusion is given in Sec. \ref{sec4}.

\section{Analysis method and observational data}
\label{sec2}

\subsection{Analysis method}

In this paper, we consider the simplest dynamical dark energy models, which means that the models we choose are those having one more parameter compared to $\Lambda$CDM, i.e., the $w$CDM model and the HDE model. We place constraints on the sterile neutrino parameters in the two considered dynamical dark energy models, and then make a comparison with the case of $\Lambda$CDM. There are six independent cosmological parameters in the base $\Lambda$CDM model,
$${\bf P}=\{\omega_b,~\omega_c,~100\theta_{\rm MC},~\tau,~\ln (10^{10}A_s),~n_s\},$$
where $\omega_b\equiv \Omega_b h^2$ and $\omega_c\equiv \Omega_c h^2$ are the present-day baryon and cold dark matter densities, respectively, $\theta_{\rm MC}$ is the ratio between the sound horizon and the angular diameter distance at the decoupling epoch, $\tau$ is the Thomson scattering optical depth due to reionization, $A_s$ is the amplitude of initial curvature perturbation power spectrum at $k=0.05~{\rm Mpc}^{-1}$, and $n_s$ is its  spectral index. For dynamical dark energy models, an additional parameter $w$ needs to be considered in the $w$CDM model and an additional parameter $c$ needs to be considered in the HDE model. Thus, there are seven independent parameters in total for the $w$CDM model and the HDE model. In addition, there are two additional free parameters, $N_{\rm eff}$ and $m_{\nu,{\rm sterile}}^{\rm eff}$, for describing the sterile neutrino species (note that here $N_{\rm eff}>3.046$). In this work we assume the minimal-mass normal hierarchy for active neutrinos, i.e., we fix the total mass of three-generation active neutrinos as $\sum m_\nu=0.06$ eV. When the sterile neutrino species is considered in the $\Lambda$CDM model, the $w$CDM model, and the HDE model, these cases are thus called the $\Lambda$CDM+$\nu_s$ model, the $w$CDM+$\nu_s$ model, and the HDE+$\nu_s$ model, respectively, in this paper. Thus, the $\Lambda$CDM+$\nu_s$ model has eight base parameters, and the $w$CDM+$\nu_s$ model and the HDE+$\nu_s$  model have nine base parameters.

The latest version of Monte Carlo Markov chain package {\tt CosmoMC}  \cite{Lewis:2002ah} is employed in our calculation to infer the posterior probability distributions of parameters.
We will make a comparison for the three models (the $\Lambda$CDM+$\nu_s$ model, the $w$CDM+$\nu_s$ model, and the HDE+$\nu_s$  model) in this paper. Because there are different numbers of parameters for the three models, we employ the Akaike information criterion (AIC) \cite{AIC1974} to do the model comparison. The AIC is defined as ${\rm AIC}=\chi_{\rm min}^2+2k$,  where $k$ is the number of parameters of a model. In practice, we do not care about the absolute value of the criterion, and we actually pay more attention to the relative values between different models, i.e., $\Delta {\rm AIC}=\Delta\chi_{\rm min}^2+2\Delta k$. A model with a lower AIC value is more favored by data. In this paper, we take the $\Lambda$CDM+$\nu_s$ model as a reference model and compare the cases of dynamical dark energy with it.

\subsection{Observational data}

In this paper, we will search for sterile neutrinos in dynamical dark energy models with the current observations, and compare the constraint results with the case of $\Lambda$CDM. We consider the following data sets:

\begin{itemize}
 \item {\it The Planck data}: We use the Planck 2015 data release of CMB temperature and polarization anisotropies, denoted as ``Planck TT,TE,EE+lowP" \cite{Aghanim:2015xee}, following the Planck collaboration \cite{Ade:2015xua}.

 \item {\it The BAO data}: The BAO distance scale data are used in this paper to break the geometric degeneracy. We use the 6dFGS sample \cite{Beutler:2011hx}, the SDSS MGS sample \cite{Ross:2014qpa}, and the LOWZ and CMASS samples of BOSS DR12 \cite{Cuesta:2015mqa}.

 \item {\it The SN data}: We use the ``joint light-curve analysis'' (JLA) sample \cite{Betoule:2014frx}, compiled from the SNLS, SDSS, and the samples of several low-redshift SN data.

 \item {\it The $H_0$ measurement}: We employ the latest measurement of the Hubble constant, $H_0=73.00{\pm1.75}~{\rm km}~{\rm s}^{-1}~{\rm Mpc}^{-1}$, reported in Ref.~\cite{Riess:2016jrr}.

 \item {\it The lensing data}: We use the CMB lensing power spectrum from the Planck lensing measurement \cite{Ade:2015zua}, which provides additional information at low redshifts. Since the CMB lensing reconstruction data directly probe the lensing  power, they are sensitive to the neutrino mass.

\end{itemize}

There are actually also a number of measurements of growth of structure that are rather important for probing the properties of dark energy, active neutrinos, as well as sterile neutrinos. For example, the observations of weak gravitational lensing (WL), redshift space distortions (RSD), and galaxy cluster counts have been widely used to search for sterile neutrinos \cite{Zhao:2017jma,Zhao:2017urm,Feng:2017nss}, to weigh active neutrino mass \cite{Guo:2017hea,Zhao:2016ecj}, and to distinguish between dark energy and modified gravity \cite{Ade:2015rim}. However, currently, some significant, uncontrolled systematics still remain in these large-scale structure observations. In the paper \cite{Ade:2015xua} of Planck collaboration, they do not use RSD or galaxy weak lensing measurements for combined constraints just because of the uncontrolled systematics remained in these measurements. In this paper, to be in accordant with the Planck collaboration \cite{Ade:2015xua}, we do not use RSD, WL, or cluster counts measurements for combined constraints either. The usage of the observational data in this paper is identical to our previous studies \cite{Zhang:2015uhk,Wang:2016tsz} on the cosmological weighing of active neutrinos, which is fairly convenient for a direct comparison of these studies.

In what follows, we will use these observational data to place constraints on the dynamical dark energy cosmologies involving sterile neutrinos. We will compare the dynamical dark energy models with the $\Lambda$CDM model, according to the constraints from two data combinations, i.e., Planck TT,TE,EE+lowP+BAO and Planck TT,TE,EE+lowP+BAO+SN+$H_0$+lensing. In the next section, we will report and discuss the fitting results of the cosmological models in the light of these data sets.

\section{Results and discussion}\label{sec3}

%%%%%%%%%%%%%%%table%%%%%%%%%%%%%%%%%%%%%%%%
\begin{table*}\small
\setlength\tabcolsep{0.5pt}
\renewcommand{\arraystretch}{1.2}
\caption{Fitting results for the $\Lambda$CDM+$\nu_s$ model, the $w$CDM+$\nu_s$ model, and the HDE+$\nu_s$ model. We quote $\pm 1\sigma$ errors, but for the parameters that cannot be well constrained, we quote the 95.4\% CL upper limits.}
\centering
\begin{tabular}{ccccccccc}\hline
\hline \multicolumn{1}{c}{Data} &\multicolumn{3}{c}{Planck TT,TE,EE+lowP+BAO}&&\multicolumn{3}{c}{Planck TT,TE,EE+lowP+BAO+SN+$H_{0}$+lensing}&\\
\cline{1-1}\cline{2-4}\cline{6-8}

  Model & $\Lambda$CDM+$\nu_s$ &$w$CDM+$\nu_s$ &HDE+$\nu_s$ &&$\Lambda$CDM+$\nu_s$ &$w$CDM+$\nu_s$&HDE+$\nu_s$\\

\hline
$\Omega_bh^2$&$0.02244^{+0.00015}_{-0.00018}$&$0.02240^{+0.00016}_{-0.00019}$&$0.02246^{+0.00016}_{-0.00021}$&&$0.02257\pm0.00017$&$0.02247^{+0.00017}_{-0.00020}$&$0.02268^{+0.00021}_{-0.00022}$\\
$\Omega_ch^2$&$0.1187^{+0.0044}_{-0.0027}$&$0.1192^{+0.0039}_{-0.0025}$&$0.1189^{+0.0042}_{-0.0029}$&&$0.1209\pm0.0030$&$0.1204^{+0.0033}_{-0.0026}$&$0.1210^{+0.0033}_{-0.0030}$\\
$100\theta_{MC}$&$1.04075^{+0.00040}_{-0.00033}$&$1.04069^{+0.00038}_{-0.00035}$&$1.04076^{+0.00041}_{-0.00035}$&&$1.04066^{+0.00043}_{-0.00040}$&$1.04068^{+0.00040}_{-0.00036}$&$1.04066^{+0.00041}_{-0.00040}$\\
$\tau$&$0.090\pm0.017$&$0.086\pm0.018$&$0.095\pm0.017$&&$0.080^{+0.014}_{-0.013}$&$0.073\pm0.015$&$0.098\pm0.014$\\
${\rm{ln}}(10^{10}A_s)$&$3.116^{+0.034}_{-0.035}$&$3.110\pm0.036$&$3.126\pm0.035$&&$3.099^{+0.026}_{-0.029}$&$3.084^{+0.030}_{-0.032}$&$3.134\pm0.029$\\
$n_s$&$0.9719^{+0.0050}_{-0.0078}$&$0.9695^{+0.0054}_{-0.0078}$&$0.9734^{+0.0060}_{-0.0087}$&&$0.9782^{+0.0067}_{-0.0076}$&$0.9732^{+0.0065}_{-0.0086}$&$0.9834^{+0.0084}_{-0.0086}$\\
\hline
$m_{\nu,{\rm{sterile}}}^{\rm{eff}}\,[\rm eV]$&$<0.7279$&$<0.7034$&$<0.6475$&&$<0.2675$&$<0.5313$&$<0.1989$\\
$N_{\rm eff}$&$<3.4273$&$<3.3941$&$<3.4674$&&$<3.5718$&$<3.5008$&$<3.6701$\\
\hline
$w$&...&$-1.055^{+0.073}_{-0.060}$&$...$&&...&$-1.058\pm0.044$&$...$\\
$c$&...&$...$&$0.541^{+0.043}_{-0.054}$&&...&$...$&$0.628^{+0.034}_{-0.041}$\\
$\Omega_m$&$0.3085^{+0.0067}_{-0.0068}$&$0.300\pm0.012$&$0.273\pm0.014$&&$0.3018\pm0.0065$&$0.2957^{+0.0081}_{-0.0079}$&$0.2898^{+0.0077}_{-0.0078}$\\
$\sigma_8$&$0.817^{+0.030}_{-0.021}$&$0.825^{+0.031}_{-0.025}$&$0.849^{+0.032}_{-0.027}$&&$0.816^{+0.019}_{-0.013}$&$0.816^{+0.024}_{-0.015}$&$0.828^{+0.017}_{-0.013}$\\
$H_0\,[{\rm km}/{\rm s}/{\rm Mpc}]$&$68.30^{+0.53}_{-1.00}$&$69.50^{+1.50}_{-1.80}$&$72.70^{+1.90}_{-2.20}$&&$69.35^{+0.83}_{-1.19}$&$70.10\pm1.00$&$70.70\pm1.10$\\
\hline
$\chi^2_{\rm min}$&12953.554&12952.890&12966.166&&13677.858&13675.836&13692.702\\
\hline
\hline
\end{tabular}
\label{tab1}
\end{table*}
%%%%%%%%%%%%%%%%%%%%%%%%%%%%%%%%%%%%%%%%%%%%%%%%
%%%%%fig1
\begin{figure*}[!htp]
\includegraphics[scale=0.32]{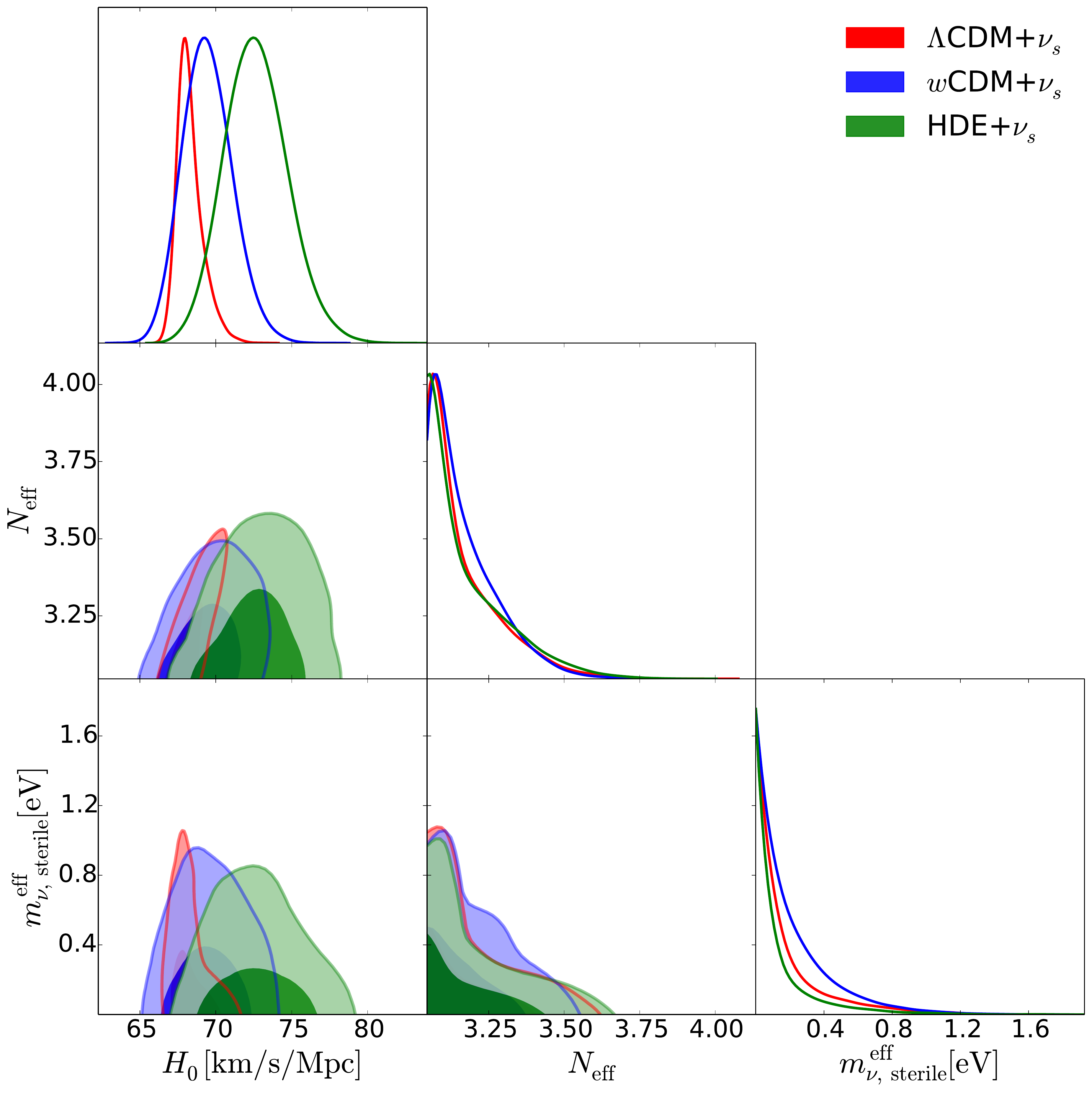}
\centering
\caption{\label{fig1} The one-dimensional posterior distributions and two-dimensional marginalized contours (68.3\% and 95.4\% CL) for the $\Lambda$CDM+$\nu_s$ ({\it red}), $w$CDM+$\nu_s$ ({\it blue}), and HDE+$\nu_s$ ({\it green}) models, from the constraints of the Planck TT,TE,EE+lowP+BAO data combination.}
\end{figure*}

%%%%%%%%%%%%%%%%%%%%%%%%%%%%%%%%%%%%%%%%%%%%%
%%%%%fig2
\begin{figure*}[!htp]
\includegraphics[scale=0.32]{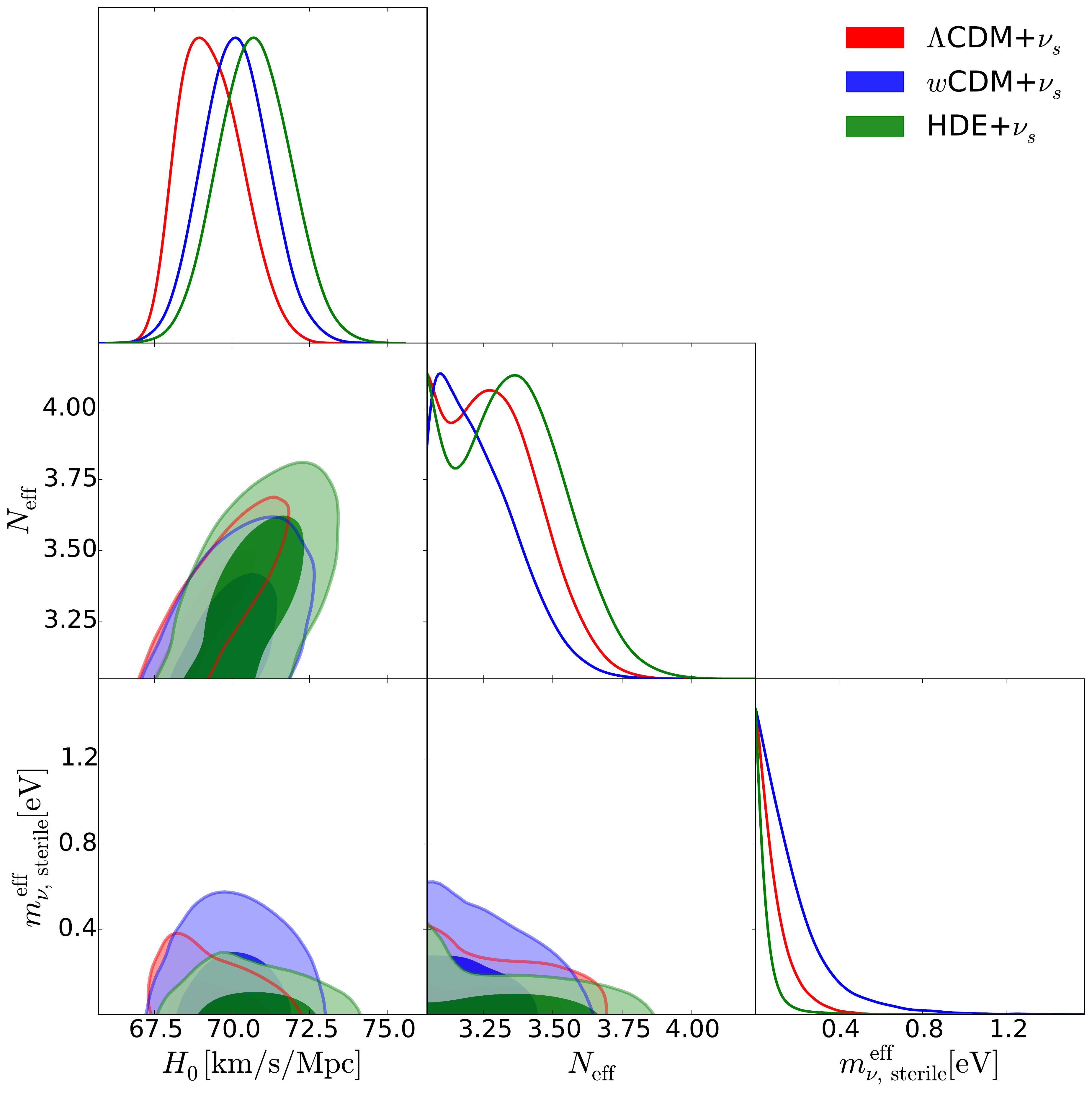}
\centering
\caption{\label{fig2} The one-dimensional posterior distributions and two-dimensional marginalized contours (68.3\% and 95.4\% CL) for the $\Lambda$CDM+$\nu_s$ ({\it red}), $w$CDM+$\nu_s$ ({\it blue}), and HDE+$\nu_s$ ({\it green}) models, from the constraints of the Planck TT,TE,EE+lowP+BAO+SN+$H_{0}$+lensing data combination.}
\end{figure*}

%%%%%%%%%%%%%%%%%%%%%%%%%%%%%%%%%%%%%%%%%%%%%%
%%%%%fig3
\begin{figure*}[!htp]
\includegraphics[scale=0.9]{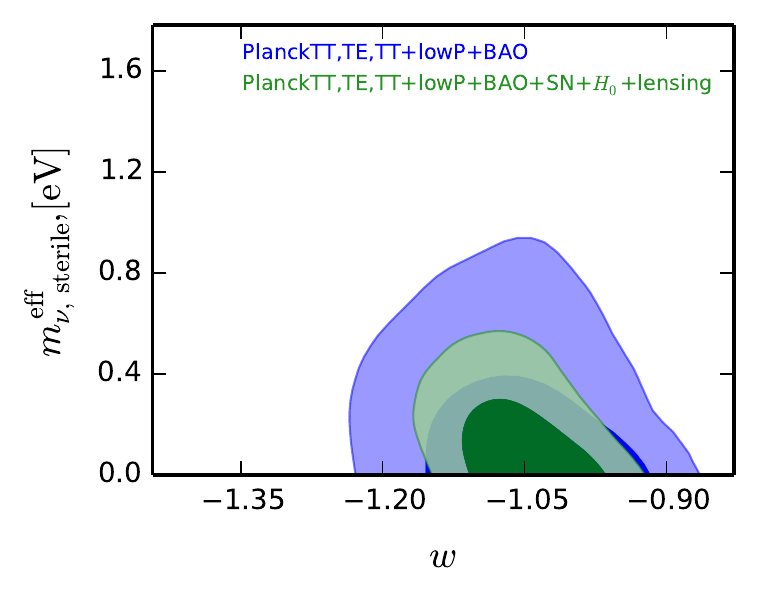}
\includegraphics[scale=0.9]{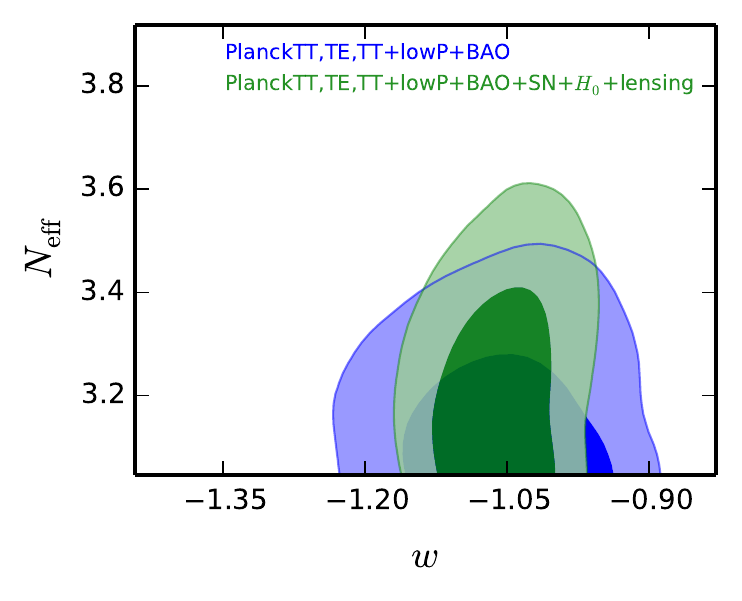}
\centering
 \caption{\label{fig3} Constraint results for the  $w$CDM+$\nu_s$ model from the Planck TT,TE,EE+lowP+BAO data combination and the Planck TT,TE,EE+lowP+BAO+SN+$H_{0}$+lensing data combination. Left panel: two-dimensional marginalized posterior contours (68.3\% and 95.4\% CL) in the $m_{\nu,{\rm{sterile}}}^{\rm{eff}}$--$w$ plane. Right panel: two-dimensional marginalized posterior contours (68.3\% and 95.4\% CL) in the $N_{\rm{eff}}$--$w$ plane.}
\end{figure*}

%%%%%%%%%%%%%%%%%%%%%%%%%%%%%%%%%%%%%%%%%%%%%%
%%%%%fig4
\begin{figure*}[!htp]
\includegraphics[scale=0.9]{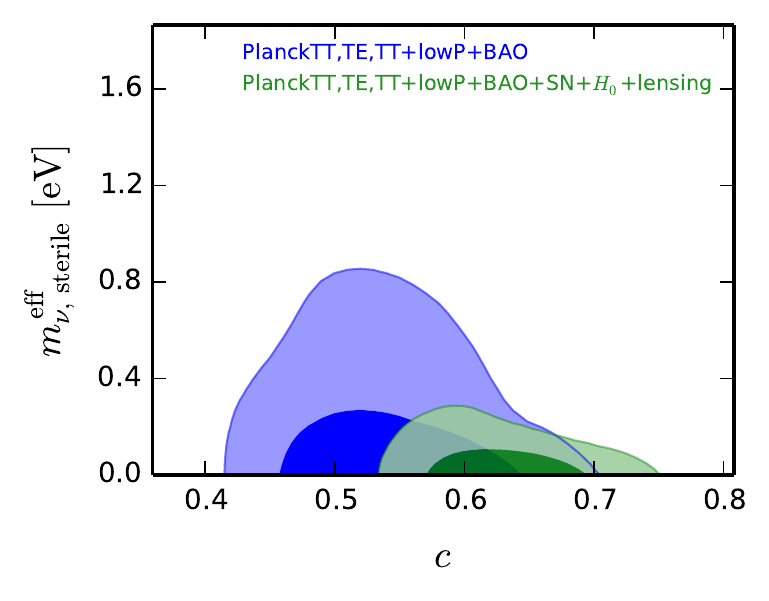}
\includegraphics[scale=0.9]{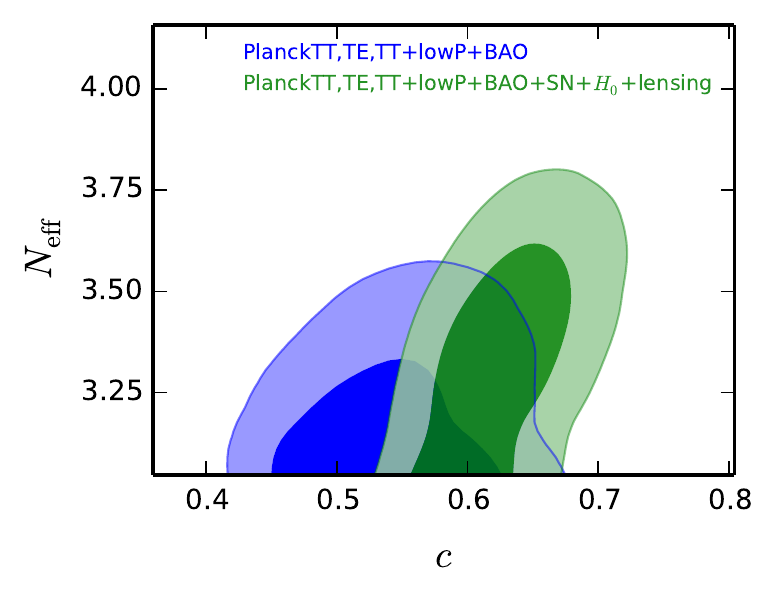}
\centering
 \caption{\label{fig4} Constraint results for the HDE+$\nu_s$ model from the Planck TT,TE,EE+lowP+BAO data combination and the Planck TT,TE,EE+lowP+BAO+SN+$H_{0}$+lensing data combination. Left panel: two-dimensional marginalized posterior contours (68.3\% and 95.4\% CL) in the $m_{\nu,{\rm{sterile}}}^{\rm{eff}}$--$c$ plane. Right panel: two-dimensional marginalized posterior contours (68.3\% and 95.4\% CL) in the $N_{\rm{eff}}$--$c$ plane.}
\end{figure*}

%%%%%%%%%%%%%%%%%%%%%%%%%%%%%%%%%%%%%%%%%%%%%%%%
%%%%%fig5
\begin{figure*}[!htp]
\includegraphics[scale=0.8]{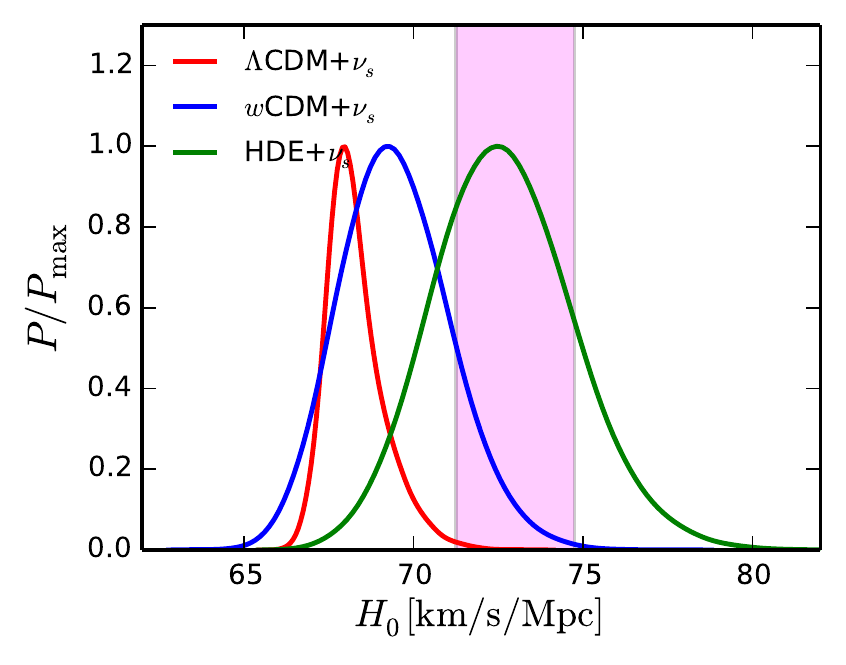}
\includegraphics[scale=0.8]{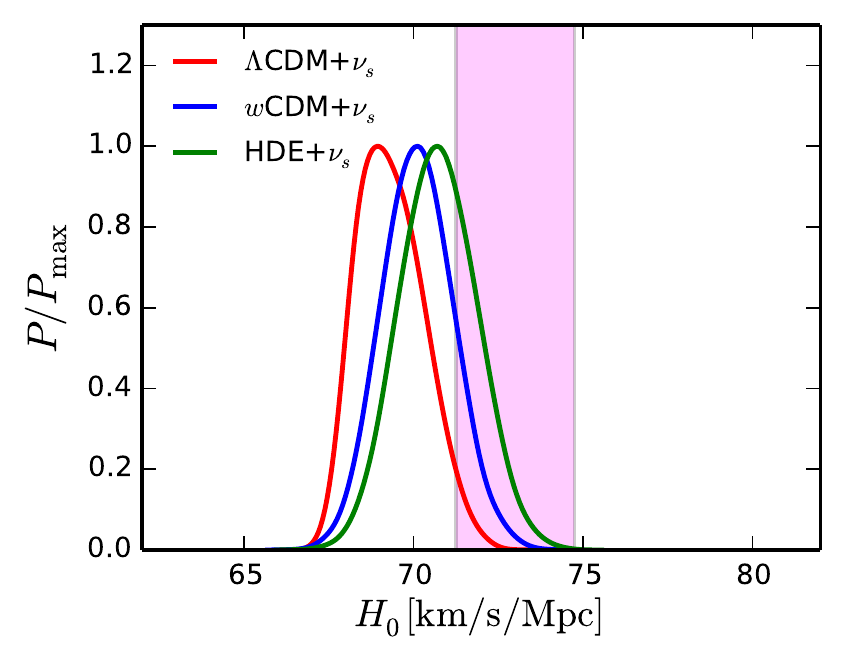}
\centering
 \caption{\label{fig5} The one-dimensional posterior distributions of $H_0$ in the $\Lambda$CDM+$\nu_s$, $w$CDM+$\nu_s$, and HDE+$\nu_s$ models, from the constraints of Planck TT,TE,EE+lowP+BAO ({\it left}) and Planck TT,TE,EE+lowP+BAO+SN+$H_0$+lensing ({\it right}). }
\end{figure*}

%%%%%%%%%%%%%%%%%%%%%%%%%%%%%%%%%%%%%%%%%%%%%%

In this section, we report the fitting results of the three models ($\Lambda$CDM+$\nu_s$, $w$CDM+$\nu_s$, and HDE+$\nu_s$) and discuss the impacts of dark energy on constraining the sterile neutrino parameters.

The fitting results are shown in Table \ref{tab1} and Figs.~\ref{fig1}--\ref{fig5}. In Table \ref{tab1}, we quote the $\pm1\sigma$ errors, but for the parameters that cannot be well constrained, we quote the 95.4\% CL upper limits.

In the case of Planck TT,TE,EE+lowP+BAO constraints, we have $m_{\nu,{\rm{sterile}}}^{\rm{eff}}<0.7279$ eV and $N_{\rm eff}<3.4273$ for $\Lambda$CDM+$\nu_s$, $m_{\nu,{\rm{sterile}}}^{\rm{eff}}<0.7034$ eV and $N_{\rm eff}<3.3941$ for $w$CDM+$\nu_s$, and $m_{\nu,{\rm{sterile}}}^{\rm{eff}}<0.6475$ eV and $N_{\rm eff}<3.4674$ for HDE+$\nu_s$. We find that the fit results of $m_{\nu,{\rm{sterile}}}^{\rm{eff}}$ and $N_{\rm eff}$ are similar for the three models. This is because the data combination Planck TT,TE,EE+lowP+BAO does not sensitive to the parameters of sterile neutrino and they actually cannot give a tight constraint on the sterile neutrino parameters.

In Fig.~\ref{fig1}, we compare the constraint results of the three models in the fit to the Planck TT,TE,EE+lowP+BAO data combination. We find that $N_{\rm eff}$ is positively correlated with $H_0$ for all the three models, thus considering the parameter $N_{\rm eff}$ in cosmological models can indeed help enhance the fit value of $H_0$. We can clearly see that, for both the $w$CDM+$\nu_s$ and HDE+$\nu_s$ models, the posterior distributions of $H_0$ become much wider compared to the $\Lambda$CDM+$\nu_s$ case, and for the HDE+$\nu_s$ model, the $H_0$ distribution gets a considerable right shift. But for the distributions of $m_{\nu,{\rm{sterile}}}^{\rm{eff}}$ and $N_{\rm eff}$, the differences for the three models are not evident.

%but the results of $H_0$ are quite different. This indicates that the dark energy properties play an important role in changing the fit results.

The BAO distance scales measurements are considered to break the geometric degeneracy, but not enough, thus adding the low-redshift observations can help further break the parameter degeneracy. In the case of the Planck TT,TE,EE+lowP+BAO+SN+$H_0$+lensing, we have $m_{\nu,{\rm{sterile}}}^{\rm{eff}}<0.2675$ eV and $N_{\rm eff}<3.5718$ for $\Lambda$CDM+$\nu_s$, $m_{\nu,{\rm{sterile}}}^{\rm{eff}}<0.5313$ eV and $N_{\rm eff}<3.5008$ for $w$CDM+$\nu_s$, and $m_{\nu,{\rm{sterile}}}^{\rm{eff}}<0.1989$ eV and $N_{\rm eff}<3.6701$ for HDE+$\nu_s$. Obviously, we see that adding the low-redshift data strikingly changes the values of the upper limit of $m_{\nu,{\rm{sterile}}}^{\rm{eff}}$.

Including low-redshift data tightens the constraints on $m_{\nu,{\rm{sterile}}}^{\rm{eff}}$ for the three models.
For the $w$CDM+$\nu_s$ model, the constraint is weakest, and for the HDE+$\nu_s$ model, the constraint is most stringent, which is in accordance with the conclusions in previous studies for the active neutrino mass \cite{Zhang:2015uhk} (see also Ref.~\cite{Wang:2016tsz}), i.e., it was found that the HDE model leads to the extremely stringent limit on the total mass of active neutrinos ($\sum m_\nu<0.113$ eV), with the data combination Planck TT,TE,EE+lowP+BAO+SN+$H_0$+lensing. Similarly, with regard to the sterile neutrino species, the HDE cosmology also leads to an extremely stringent upper limit on the effective mass of sterile neutrino ($m_{\nu,{\rm{sterile}}}^{\rm{eff}}<0.1989$ eV).

Besides, for the constraint results for the parameter $N_{\rm eff}$, we find that $N_{\rm eff}$ cannot be well constrained using the Planck TT,TE,EE+lowP+BAO+SN+$H_0$+lensing data, but the addition of the measurement of growth of structure data can significantly improve the constraint on $N_{\rm eff}$. For example, in Ref.~\cite{Feng:2017nss} the detection of $\Delta N_{\rm eff}>0$ in the $\Lambda$CDM cosmology is at the 1.27$\sigma$ level with the data combination of Planck TT,TE,EE+lowP+BAO+SZ+$H_0$+lensing+WL, and in Ref. \cite{Zhao:2017urm} the detection of $\Delta N_{\rm eff}>0$ in the holographic dark energy cosmology is at the 2.75$\sigma$ level with the data combination of Planck TT+lowP+BAO+SN+RSD+WL+lensing+$H_0$ (here, SZ refers to Planck thermal Sunyaev-Zeldovich clusters measurement, WL refers to weak lensing galaxy shear measurement, and RSD refers to redshift space distortions measurement). In fact, we only consider to give conservative constraints on the sterile neutrino species, and so we do not use the measurements of growth of structure for combined constraints in this paper.

Figure~\ref{fig2} shows the comparison of the three models for the constraints from Planck TT,TE,EE+lowP+BAO+SN+$H_0$+lensing. We can clearly see that, in this case, $N_{\rm eff}$ is still positively correlated with $H_0$, and in the dynamical dark energy models the distributions of $H_0$ get an explicit right shift. Comparing the three models, we find that the HDE+$\nu_s$ model acquires the highest $H_0$ and $N_{\rm eff}$ values and the lowest $m_{\nu,{\rm{sterile}}}^{\rm{eff}}$ value. In the $w$CDM+$\nu_s$ model, the fit value of $m_{\nu,{\rm{sterile}}}^{\rm{eff}}$ is highest among the models.

%Therefore, for the dark energy model with the massive sterile neutrino, the observations give a upper limit on $m_{\nu,{\rm{sterile}}}^{\rm{eff}}$ and $N_{\rm eff}$, which implies that the current observations cannot favor a massive sterile neutrino.

To directly show how dark energy properties affect the constraints on sterile neutrino parameters, the dependence of $m_{\nu,{\rm{sterile}}}^{\rm{eff}}$ and $N_{\rm eff}$ upon the parameters of dynamical dark energy models is plotted in Figs.~\ref{fig3} and \ref{fig4}. Both figures are obtained by using the two data combinations.

For the $w$CDM+$\nu_s$ model, from Table \ref{tab1}, we have $w=-1.055^{+0.073}_{-0.060}$ without the low-redshift data, and it changes to $w=-1.058{\pm0.044}$ after considering the low-redshift data. Obviously, adding the low-redshift data tightens the constraints on the dark energy parameter $w$. In Fig.~\ref{fig3}, we find that $m_{\nu,{\rm{sterile}}}^{\rm{eff}}$ is evidently anti-correlated with $w$. And, adding the low-redshift data significantly changes the limits of  $m_{\nu,{\rm{sterile}}}^{\rm{eff}}$ and $N_{\rm eff}$: the upper limit on $m_{\nu,{\rm{sterile}}}^{\rm{eff}}$ is driven to a lower value, while the upper limit on $N_{\rm eff}$ is driven to a higher value.

For the HDE+$\nu_s$ model, from Table \ref{tab1}, we have $c=0.541^{+0.043}_{-0.054}$ without the low-redshift data, and it changes to $c=0.628^{+0.034}_{-0.041}$ with the addition of the low-redshift data. We find that the Planck CMB data prefer a low value of $c$, and the low-redshift observations tend to drive $c$ upward to a higher value. The two-dimensional posterior distribution contours (68.3\% and 95.4\% CL) in the $m_{\nu,{\rm{sterile}}}^{\rm{eff}}$--$c$ plane and the $N_{\rm eff}$--$c$ plane, for the constraints from two data combinations, are shown in Fig.~\ref{fig4}. We find that $m_{\nu,{\rm{sterile}}}^{\rm{eff}}$ is slightly anti-correlated with $c$. Evidently, adding the low-redshift data tightens the constraints on the dark energy parameter $c$ and significantly changes the limits of $m_{\nu,{\rm{sterile}}}^{\rm{eff}}$ and $N_{\rm eff}$. As the same to the case of $w$CDM, including the low-redshift data drives $m_{\nu,{\rm{sterile}}}^{\rm{eff}}$ to a lower value and drives $N_{\rm eff}$ to a higher value for their upper limits.

%Therefore, we conclude that the dark energy properties can significant impact the constraint limits on sterile neutrino parameters.

We now make a simple model comparison for the three models. The value of $\chi^2_{\rm min}$ for different models in the fit are also listed in Table~\ref{tab1}. For the $w$CDM+$\nu_s$ model, the value of $\chi^2_{\rm min}$ is similar to the $\Lambda$CDM+$\nu_s$ model for Planck TT,TE,EE+lowP+BAO, but becomes slightly smaller after considering the low-redshift data. In the all-data case, the $w$CDM+$\nu_s$ model leads to an increase of $\Delta\chi^2=-2.022$ compared with the $\Lambda$CDM+$\nu_s$ model, indicating that in this case the $w$CDM+$\nu_s$ model can improve the fit. Actually, when we use the information criterion to make a model selection, we have $\Delta {\rm AIC}=-0.022$ for the $w$CDM+$\nu_s$ model in this case, which shows that the $w$CDM+$\nu_s$ model is only slightly better than the $\Lambda$CDM+$\nu_s$ model from the statistical point of view. For the HDE+$\nu_s$ model, the values of $\chi^2_{\rm min}$ are much larger than those of the $\Lambda$CDM+$\nu_s$ model (the reason is that the HDE model cannot fit the BAO point at $z_{\rm eff}=0.57$ and the JLA data well in the global fit). Compared to the $\Lambda$CDM+$\nu_s$ model, the HDE+$\nu_s$ model leads to an increase of $\Delta\chi^2=12.622$ for the Planck TT,TE,EE+lowP+BAO constraint and an increase of $\Delta\chi^2=14.844$ for the Planck TT,TE,EE+lowP+BAO+SN+$H_0$+lensing constraint. That is to say, the HDE+$\nu_s$ model has $\Delta {\rm AIC}=14.622$ and $\Delta {\rm AIC}=16.844$ for the two data combination cases, which shows that the HDE+$\nu_s$ model is not favored by the current cosmological observations, compared to the case of $\Lambda$CDM.

Next, we wish to simply address the tension issue. In fact, the tension between the Planck observation and the local measurement of the Hubble constant has been studied widely \cite{Sola:2017znb,DiValentino:2017iww,DiValentino:2015wba,Zhao:2017urm,Feng:2017nss,Qing-Guo:2016ykt,DiValentino:2016hlg,Dvorkin:2014lea,Zhang:2014dxk,Battye:2013xqa,Hamann:2013iba,Wyman:2013lza,Li:2013dha,Zhang:2017epd,Guo:2017qjt}. To reduce the tension, more possibilities should be explored. One interesting suggestion is to consider additional sterile neutrino species in the universe \cite{Feng:2017nss,Dvorkin:2014lea,Zhang:2014dxk,Battye:2013xqa,Hamann:2013iba,Wyman:2013lza}, and another interesting suggestion is to consider the dynamical dark energy cosmology \cite{DiValentino:2016hlg,DiValentino:2015wba,Li:2013dha,Qing-Guo:2016ykt}. Of course, simultaneously consider the two aspects would furthest relieve the tension.

%Furthermore, in order to reduce the tension, we consider the massive sterile neutrinos in dynamical dark energy cosmology in this paper.

In Fig.~\ref{fig5}, we show the one-dimensional posterior distributions of $H_0$ for the three models. In the left panel, we show the case of fitting to the Planck TT,TE,EE+lowP+BAO data, and in the right panel, we show the case of fitting to the Planck TT,TE,EE+lowP+BAO+SN+$H_0$ +lensing data. From this figure, we can clearly see that once the sterile neutrinos are considered in the dynamical dark energy models, then according to the constraints of the two data combinations, the fit value of $H_0$ can become much larger. In Table~\ref{tab1}, for the constraints of Planck TT,TE,EE+lowP+BAO data, we obtain $H_0=68.30^{+0.53}_{-1.00}~{\rm km}~{\rm s}^{-1}~{\rm Mpc}^{-1}$ for the $\Lambda$CDM+$\nu_s$ model, $H_0=69.50^{+1.50}_{-1.80}~{\rm km}~{\rm s}^{-1}~{\rm Mpc}^{-1}$ for the $w$CDM+$\nu_s$ model, and $H_0=72.70^{+1.90}_{-2.20}~{\rm km}~{\rm s}^{-1}~{\rm Mpc}^{-1}$ for the HDE+$\nu_s$ model. This indicates that the tension with the local value of the Hubble constant, $H_0=73.00{\pm1.75}~{\rm km}~{\rm s}^{-1}~{\rm Mpc}^{-1}$, is at 2.57$\sigma$ level, 1.52$\sigma$ level, and  0.12$\sigma$ level, respectively. We find that in the $w$CDM+$\nu_s$ model, the tension is slightly relieved compared with the $\Lambda$CDM+$\nu_s$ model, while in the HDE+$\nu_s$ model the tension is largely relieved. For the constraints of Planck TT,TE,EE+lowP+BAO +SN+$H_0$+lensing data, we obtain $H_0=69.35^{+0.83}_{-1.19}~{\rm km}~{\rm s}^{-1}~{\rm Mpc}^{-1}$ for the $\Lambda$CDM+$\nu_s$ model, $H_0=70.10{\pm1.00}~{\rm km}~{\rm s}^{-1}~{\rm Mpc}^{-1}$ for the $w$CDM+$\nu_s$ model, and $H_0=70.70{\pm1.10}~{\rm km}~{\rm s}^{-1}~{\rm Mpc}^{-1}$ for the HDE+$\nu_s$ model. In this case, the tension with the Hubble constant direct measurement is at 1.88$\sigma$ level, 1.44$\sigma$ level, and 1.11$\sigma$ level, respectively. Thus we have shown that the tension can be largely relieved once sterile neutrinos and dynamical dark energy are both considered. Therefore, compared to the $\Lambda$CDM+$\nu_s$ model, the dynamical dark energy models with sterile neutrinos can indeed largely relieve the tension between the Planck observation and the local measurement of the Hubble constant. In this aspect, the HDE model performs best. But, recalling the analysis above, it should be reminded that the HDE+$\nu_s$ model is evidently worse than the $\Lambda$CDM+$\nu_s$ model in the fit to the current observations. Therefore, the scheme to solve the $H_0$ tension by considering the HDE+$\nu_s$ model is actually not strongly supported by the observations.

\section{Conclusion}\label{sec4}

We constrain the sterile neutrino parameters, $m_{\nu,{\rm{sterile}}}^{\rm{eff}}$ and $N_{\rm eff}$, in dynamical dark energy models. We only consider the simplest extensions to the $\Lambda$CDM cosmology, namely, those having only one more parameter than $\Lambda$CDM. Thus, in this paper, we search for sterile neutrinos in the $w$CDM and HDE cosmologies. We use the latest observational data sets (including the Planck 2015 TT,TE,EE+lowP, BAO, SN, $H_0$, and Planck lensing measurements) to do the analysis.

We have the following findings: (i) Without the addition of the measurements of growth of structure, only upper limits of $m_{\nu,{\rm{sterile}}}^{\rm{eff}}$ and $N_{\rm eff}$ can be obtained, which indicates that no evidence of the existence of a sterile neutrino species with eV-scale mass is found in this analysis. (ii) Dark energy properties could significantly influence the constraint limits of sterile neutrino parameters. Compared to the $\Lambda$CDM cosmology, in the $w$CDM cosmology the limit of $m_{\nu,{\rm{sterile}}}^{\rm{eff}}$ becomes much looser, but in the HDE cosmology the limit becomes much tighter. (iii) In the cosmological fit, the $w$CDM+$\nu_s$ model is only slightly better than the $\Lambda$CDM+$\nu_s$ model, and the HDE+$\nu_s$ model is much worse than the $\Lambda$CDM+$\nu_s$ model. (iv) In the HDE+$\nu_s$ model, the $H_0$ tension is indeed furthest relieved, but this does not mean that the problem is really solved because the HDE+$\nu_s$ model is not favored by the current observations from a model comparison analysis.

\begin{acknowledgments}

This work was supported by the National Natural Science Foundation of China (Grants No.~11522540 and No.~11690021), the National Program for Support of Top-notch Young Professionals, and the Provincial Department of Education of Liaoning (Grant No.~L2012087).

\end{acknowledgments}

%\paragraph{Note added.} This is also a good position for notes added
%after the paper has been written.


\end{document}